\documentclass[useAMS,usenatbib,graphicx]{mn2e}
\usepackage[totalwidth=480pt, totalheight=680pt]{geometry}
\usepackage{graphicx}
\usepackage{amssymb}
\usepackage{amsbsy}   
\usepackage{amsmath}

\newcommand{\ms}{$M_{stellar}$~}
\newcommand{\msun}{$M_{\odot}$}

\title[Early constraints from metal-rich GCs]{The first gigayear of bulge star formation in Virgo ellipticals:  constraints from their globular cluster systems}
\author[Lee R. Spitler]{Lee R. Spitler$^{1}$\thanks{E-mail: lspitler@astro.swin.edu.au } \\
$^{1}$Centre for Astrophysics \& Supercomputing, Swinburne University, Hawthorn, VIC 3122, Australia}

\begin{document}

\pagerange{\pageref{firstpage}--\pageref{lastpage}} \pubyear{2010}
\maketitle

\label{firstpage}

\begin{abstract}
Data products from the Advanced Camera for Surveys Virgo Cluster Survey are used to understand the bulge star formation history in early-type galaxies at redshifts $z$~$\ga$~$2$.  A new technique is developed whereby observed high-redshift age-metallicity relationships are utilized to constrain the typical formation epochs of metal-rich or ``bulge'' globular clusters.  This analysis supports a model where massive Virgo galaxies underwent an extremely intense mode of bulge globular cluster formation at $z\sim3.5$ that was followed by an era of significant bulge growth and little globular cluster production.  Intermediate-mass galaxies showed a less-intense period of globular cluster formation at $z\sim2.5$ that was synchronized with the bulk of bulge star growth.  The transition between the massive and intermediate-mass galaxy star formation modes occurs at a galaxy stellar mass of \ms $\sim3\times10^{10}$ \msun, the mass where many other galaxy properties are observed to change.  Dwarf early-type galaxies in Virgo may have experienced no significant period of bulge globular cluster formation, thus the intense star bursts associated with globular cluster formation may be difficult to directly observe at redshifts $z\la4$.  Though the above conclusions are preliminary because they are based upon uncertain relationships between age and metallicity, the technique employed will yield more stringent constraints as high-redshift galaxy observations and theoretical models improve.
\end{abstract}


\section{Introduction}\label{chapxintro}

Globular star cluster formation only occurs during strong star formation events \citep{harris_supergiant_1994,elmegreen_universal_1997,larsen_mass_2009}.  Most GCs found in early-type galaxies formed at redshifts $z\ga2$ \citep[see references in ][]{brodie_extragalactic_2006}, thus globular cluster (GC) systems are valuable observational tools to help understand the nature of major star formation events in early Universe \citep{ashman_formation_1992,forbes_origin_1997,ct_formation_1998,harris_globular_2001,brodie_extragalactic_2006}.

High-redshift galaxy observations continue to provide more and more detailed galaxy properties (e.g. ages, metallicities, sizes, star-formation rates) at the redshifts of GCs formation \citep[e.g. ][]{hopkins_normalization_2006,bouwens_uv_2007,reddy_multiwavelength_2008,franx_significant_2003,van_dokkum_spectroscopic_2003,glazebrook_high_2004,daddi_passively_2005,cimatti_gmass_2008,maiolino_amaze._2008,van_dokkum_growth_2010}. Though much progress has been made in this field, interpretations are generally limited due to the challenging nature of such observations.  By combining constraints from near-field GC system observations with far-field galaxy observations, certain limitations can be overcome and unique predictions for this important period of galaxy formation can be made \citep[e.g. ][]{shapiro_star-forming_2010}.

One of the strongest predictions resulting from GC system work is that many galaxies experienced two modes or epochs of intense star formation sometime before $z\sim2$.  This follows from observations that early-type galaxies with stellar masses of \ms~$\ga$~$10^{10}$~\msun~generally host two GC metallicity subpopulations \citep[e.g. ][]{strader_globular_2006,peng_acs_2006}.  The metal-poor or ``halo'' GC subpopulation is thought to have formed within the numerous metal-poor proto-galaxies that likely dominated the early Universe.  The metal-rich or ``bulge'' GC subpopulation likely formed later and shows similar properties to the host galaxy's bulge \citep[e.g.~metallicities, spatial distributions][]{harris_globular_2001,forbes_connection_2001,harris_halo_2002,dirsch_wide-field_2005,bassino_large-scale_2006,forte_quantitative_2007,forte_globular_2009}.

Thus by directly comparing the properties of a GC subpopulation to the galactic component that they are associated with, the relative contribution of e.g. bulge stellar mass produced during the intense MR GC formation events can be understood.  Furthermore, this type of comparison can constrain more general galaxy formation models by also considering observational and theoretical analysis of high-redshift galaxies.

The present work conducts such an analysis and compares properties of bulge GC subpopulations to their host galaxy's bulge stars.  A model of bulge star formation is developed for redshifts $z\ga2$, which is compared to existing constraints from direct, high-redshift galaxy observations.  The GC system observations come from a imaging survey of 100 early-type galaxies in the nearby ($\sim15$Mpc) Virgo Galaxy Cluster using the Advance Camera for Surveys mounted on the Hubble Space Telescope (HST).  The sample size, its homogeneity and broad range of high-level data product resulting from the Advance Camera for Surveys Virgo Cluster Survey \citep[ACSVCS; ][]{ct_acs_2004} allow for a detailed investigation into the relationship between galaxies and their GC systems, hence constraints on the first few gigayears of bulge star formation in Virgo early-type galaxies.

\subsection{Globular Cluster System Observations}\label{review}

Spectroscopy work to age-date individual GCs in elliptical galaxies has shown that the bulk of GCs are very old, with typical ages of $\ga 10$ Gyrs or $z\ga2$ \citep[e.g. ][]{larsen_keck_2002,strader_extragalactic_2005,puzia_vlt_2005,beasley_globular_2006,conselice_keck_2006,sharina_ages_2006,pierce_gemini/gmos_2006,pierce_gemini/gmos_2006-1,cenarro_stellar_2007,beasley_2df_2008}.  Because GCs are dominated by old stellar populations, the colour distribution of a galaxy's GC system can be interpreted as its intrinsic metallicity distribution \citep[see ][]{strader_globular_2007,kundu_bimodal_2007,spitler_extendingbaseline_2008}.  Observations show that most extragalactic GC systems have a bimodal colour distribution \citep{strader_globular_2006,peng_acs_2006}, which implies they host two {\it metallicity} subpopulations:  metal-poor (MP) GC subpopulations with typical metallicities of [m/H] $\sim-2$ to $-1$ and metal-rich (MR) GC subpopulations ranging [m/H] $\sim-1$ to $0$.  

MR GCs show roughly similar chemical properties to the their host galaxy's bulge \ms \citep[e.g. ][]{harris_globular_2001,forbes_connection_2001,harris_halo_2002,forte_quantitative_2007,forte_globular_2009} and the spatial distributions of MR GCs resemble the host galaxy's stellar distribution \citep[e.g. ][]{bassino_large-scale_2006}, though the GCs generally show a shallower distribution in the central regions of the galaxy possibly due to GC destruction \citep[e.g. ][]{dirsch_wide-field_2005}.  These observations suggest the formation of MR GCs and galaxy bulges are closely linked.  

MP GCs show metallicities more resembling galaxy halo stars \citep[e.g. ][]{searle_compositions_1978} and are thus thought to have formed during a ``pre-galactic'' era before galaxies started to assembly in a $\Lambda$CDM universe. Because the intrinsic metallicity distribution of early-type galaxy stellar metal-poor halos have only been studied in a few galaxies \citep[e.g. ][]{elson_red_1997,harris_halo_2002,harris_leo_2007,norris_gemini/GMOS_2008,weijmans_stellar_2009,foster_metallicity_2009}, the relationship between MP GCs and their host galaxy is more difficult to discern and hence constraints from MP GC subpopulations will not be considered here.

\section{ACSVCS Data Sample}\label{gcdata}

\begin{figure}
\center
\includegraphics[scale=0.4]{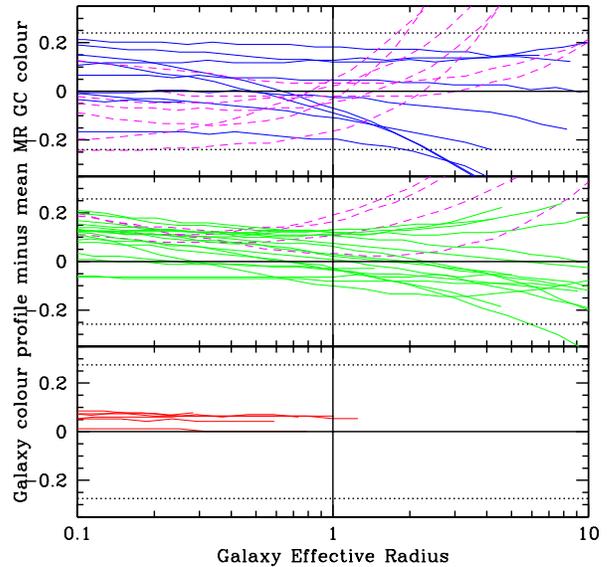}
\caption[$(g-z)_0$ colour profiles of Virgo early-type galaxies]{$(g-z)_0$ colour profiles of Virgo early-type galaxies from the S{\'e}rsic fits of \citet{ferrarese_acs_2006}.  For each galaxy, its mean MR GC colour is subtracted from the galaxy's colour profile to illustrate the relative enrichment of these two components.  Top, middle and bottom panels correspond to galaxies in the ACS Virgo Cluster Survey \citep{ct_acs_2004} with $M_{stellar} \le 10^{10} M_{\odot}$ (hereafter ``low-mass'' galaxies), $10^{10} < M_{stellar} \le 10^{11} M_{\odot}$ (``intermediate'') and $M_{stellar} > 10^{11} M_{\odot}$ (``massive''), respectively.  Colour profiles with magenta dashed lines have central blue (young) cores and are removed from further analysis (see \S\ref{gcdata}). The $\pm1\sigma$ colour boundaries for a typical MR GC subpopulation in that galaxy mass bin are represented by the two dotted lines.  All massive galaxies show colours that are redder (i.e. more enriched in metals) than the typical MR GC.  Intermediate and low-mass galaxies show a range of differences.}\label{figcolourprofile}
\endcenter
\end{figure}

The properties of MR GCs, associated with Virgo Galaxy Cluster early-type galaxies, are investigated in the following sections.  A Virgo GC system is considered to host a MR GC subpopulation if the GC system's colour (hence metallicity) distribution yields can be well-represented by two Gaussians \citep[see ][]{peng_acs_2006} and the final MR GC number estimate must not be equivalent to zero within the measurement and contamination errors.  This information is contained within \citet{peng_acs_2006} and \citet{peng_acs_2008}.  To prevent contamination from nearby GC systems, galaxies are not considered if are embedded into the projected GC system of a larger galaxy \citep[see ][]{peng_acs_2008}.  GCs are mostly old (see \S\ref{chapxintro}) and contain little dust \citep[e.g. ][]{barmby_spitzer_2009}, thus their colours are largely determined by their intrinsic metallicities.  The ACSVCS GC colours ({\it F475W}$-${\it F850LP}; hereafter $g-z$) are converted into a metallicities estimate using the empirical relation given as eqn.~2 in \citet{peng_acs_2006}.  This is valid for an old stellar population and is based upon direct observations of Milky Way and M87 GCs.  Mean GC $g-z$ colours and measured intrinsic colour dispersions are from \citet{peng_acs_2006}.

Mean galaxy metallicities of the ACSVCS sample are critical for much of the following analysis.  Compared to the simple stellar populations of GCs, observed galaxy light comes from multiple stellar component components and is thus more difficult to interpret.  Identifying the population of stars that formed during the same epoch of GC formation and then understanding their intrinsic metallicity content is hampered by the well-known age-metallicity degeneracy \citep[e.g. ][]{worthey_comprehensive_1994}.  Spectroscopic analysis can help break this degeneracy, but such work is generally limited to the central regions of a galaxy and are currently available for only a small subset of the ACSVCS sample.  Spectroscopic observations have shown that early-type galaxies are generally as old as GCs \citep{kuntschner_early-type_2002,thomas_epochs_2005,sanchez-blazquez_stellar_2006,koleva_formation_2009,smith_spectroscopic_2009}, though intermediate to lower-mass early-types can sometimes show younger central stellar populations \citep[e.g. ][]{geha_internal_2003,sanchez-blazquez_stellar_2006,sansom_bimodality_2008,smith_spectroscopic_2009}.  However, such observations can be influenced by a relatively small (in terms of \ms) young star burst that can briefly dominate a galaxy's central light \citep[e.g. ][]{lisker_virgo_2006,trager_stellar_2008,proctor_effects_2008}.  In any case, a subset of the low-mass elliptical population shows ages that are identical to old massive galaxies \citep[e.g. ][]{kuntschner_early-type_2002,koleva_formation_2009,mendel_anatomy_2009,tolstoy_star_2009}.

Analysing galaxy photometry is one way to understand the stellar populations of the ACSVCS galaxies in a homogeneous fashion.  Though galaxy broadband colours like $g-z$ are prone to uncertain interpretation, they do probe the global properties of a galaxy and therefore should be sensitive to spatial variations in the stellar populations \citep[e.g. ][]{koleva_formation_2009,spolaor_mass-metallicity_2009}, which are sometimes missed in spatially-unresolved spectroscopic analysis.  In order to study the ACSVCS sample in a homogeneous fashion, a galaxy's $(g-z)_0$ colour is transformed into a metallicity estimate by assuming the stellar population is dominated by stars of a similar age to GCs and using the same colour-metallicity transformation that is used for the GCs (see above).  This transformation is inappropriate for galaxies that show young stellar populations in their photometry.  Those with obvious young components (see below) are culled from the present analysis.  Despite this precautionary action, it is possible that a subset of the remaining ACSVCS sample could have luminosity-weighted ages that are younger than a typical GC.  The metallicity values derived for such galaxies would therefore be lower limits.  The reader will be reminded of this possibility when appropriate.

\citet{ferrarese_acs_2006} identify two galaxies in their sample (VCC~798/M85 and VCC~1499) that show significant young stellar components.  They also identify certain galaxies with corrupt or tidally truncated light profiles, which are also excluded here.  A few of the remaining galaxies in the \citet{ferrarese_acs_2006} work have blue cores in their colour-profile, consistent with a bright, young stellar population \citep[e.g. ][]{lisker_virgo_2006}, as shown in Fig.~\ref{figcolourprofile}.  Since a bright, young core complicates the estimation of a galaxy's \ms and average metallicity, these galaxies are excluded from the analysis.  In particular, if a galaxy's global $(g-z)_0$ colour is 0.2 magnitudes smaller than their colour at 10 effective radii (or the outermost radial bin covered by the HST ACS field) they are excluded.  Galaxy $(g-z)_0$ colour profiles and $z-$band effective radii are from \citet{ferrarese_acs_2006}.  Fig.~\ref{figcolourprofile} shows the ACSVCS galaxy colour gradients subtracted by each galaxy's mean MR GC colour taken from \citet{peng_acs_2006}.  Those with blue cores are highlighted in the Figure. 

Galaxy stellar masses are from 2MASS K-band luminosities (which is largely insensitive to metallicity effects), and a mass-to-light ratio (M/L$_K=0.86$) appropriate for an old stellar population \citep[see ][]{spitler_connection_2008,spitler_new_2009}.

\section{Chemical differences between MR GCs and bulge stars}\label{mrbulgediff}

As reviewed in Section~\ref{review}, MR GC subpopulations share some properties with their host galaxy's bulge stars.  In this section, the chemical similarities of the MR GCs and galaxy bulges are investigated.

\subsection{Analysis}

In Fig.~\ref{figcolourprofile}, the difference between the galaxy colour gradient and its mean MR GC subpopulation colour is presented for ACSVCS galaxies.  Massive galaxies show redder colours than the {\it typical} MR GCs by $\sim0.1$ mag, while lower mass galaxies show a spread of differences.  In the same Figure, the average $\pm1\sigma$ colour dispersion of the MR GCs in each \ms bin are presented as dotted lines to illustrate the range of colours characteristic of the MR subpopulations ($\sigma_{g-z}\pm0.15$, $\pm0.14$ and $\pm0.13$ mag. in order of the massive to low-mass galaxy bins).  Galaxy colours generally overlap with the observed intrinsic colour dispersion of their MR GC subpopulations, which is why previous work (see \S\ref{review}) has claimed that bulge stars and MR GCs are roughly similar in terms of colour and hence metallicity.

\begin{figure*}
\center
\includegraphics[scale=0.7]{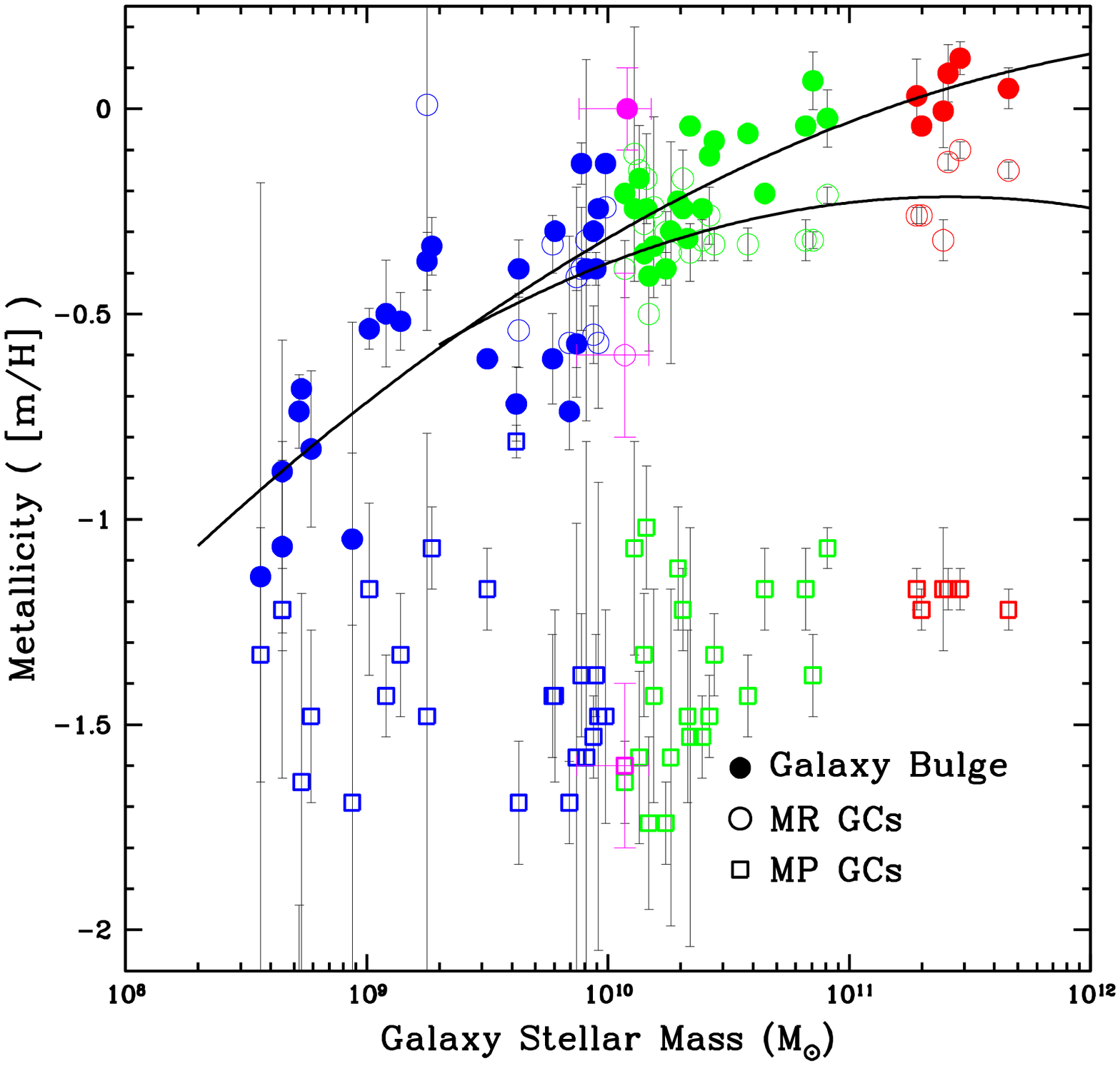}
\caption[Mean MR GC and galaxy colour against galaxy stellar mass]{Mean MR GC (open circles), MP GC (open squares) and galaxy (solid circles) metallicities against host galaxy stellar mass.  At approximately \ms $\sim10^{10}$ \msun, the MR GC and galaxy mass-metallicity relationships start to merge.  For \ms $\ga3\times10^{10}$ \ms, the two components are typically offset by a factor of $\sim2$ in metallicity.  Error bars are $1\sigma$ measurement uncertainties.  Magenta points are the Milky Way data.}\label{figcolordiff}
\endcenter
\end{figure*}

A more quantitative investigation of this relationship is presented in Fig.~\ref{figcolordiff}, where the average metallicities of both components are given.  Galaxy metallicity reflects the luminosity-weighted global colour computed from integrating the best-fit S{\'e}rsic profiles in each band to infinity and the assumption that the colours are dominated by an old stellar population (see \S\ref{gcdata}).  It is apparent in Fig.~\ref{figcolordiff} that both components become more enriched with host \ms, reflecting the well-known mass-metallicity relationship in galaxies \citep[e.g. ][]{kodama_origin_1997,gallazzi_ages_2006} and an established trend among MR GCs \citep{forbes_origin_1997,larsen_properties_2001,strader_globular_2006,peng_acs_2006}.

Also included in the Figure are MP GC mean metallicities.  The difference between the subpopulation metallicities remains relatively constant \citep[e.g. ][]{peng_acs_2006} until lower galaxy masses where MR GC subpopulations are no longer detected.  It is still possible small numbers of MR GCs are present in these galaxies, but spectroscopic confirmation of the few MR GC candidates is required.

From Fig.~\ref{figcolordiff} it is apparent that the mean metallicities of the bulge stars and MR GCs are fairly well separated for massive galaxies, but become more similar for lower mass galaxies.  For massive galaxies, the bulge stars show more enriched stellar populations than their typical MR GCs by $\sim60\%$ in metallicity.  For galaxies of \ms $\la3\times10^{10}$ \msun, the average difference between the two components apparently decreases, as the MR GC and galaxy mass-metallicity relationships begin to converge.  These galaxies have $\sim15$ MR GCs on average, so the mean MR GC metallicities should still be robust.  If a low-mass galaxy bulge metallicity is underestimated due to the presence of younger stellar populations (see \S\ref{gcdata}), then the two components of the galaxy may still be offset at lower masses.

Other systems show similar offsets in the average MR GC and bulge star metallicities. For instance, \citet{harris_halo_2002} compare the observed metallicity distributions of MR GCs and resolved bulge stars in a massive elliptical galaxy, NGC~5128, with $M_{stellar}\sim6\times10^{10}$ \msun.  The bulge stars in this galaxy at 8~kpc extend to higher metallicities than the MR GCs (see also fig.~8 in \citealt{beasley_2df_2008}), consistent with the results found here.  The enrichment level of Milky Way bulge \citep[e.g. ][]{zoccali_metal_2008} is also higher than a typical Galactic MR GC.  This is shown in Fig.~\ref{figcolordiff} using a MR GC mean metallicity from the catalogue of \citet{harris_catalog_1996} and Galactic bulge \ms from \citet{cardone_modellingmilky_2005}.

\subsection{Discussion}

A difference in the metallicity of two stellar populations can reflect a separation in their respective formation epochs if the gaseous metals used to form stars were well-mixed and the gas enriched over time.  Assuming the observed metallicity difference between galaxy bulge stars and MR GCs is a rough timescale indicator, the data in Fig.~\ref{figcolordiff} imply that the dominant bulge star-formation epoch in massive galaxies was separated in time from the peak of the MR GC formation, because the bulges are generally more enriched than the typical MR GC they host.  Since GCs require unique conditions to form, differences in the formation epochs of the MR GCs and bulge stars likely reflects a change in the star-formation mode of the host galaxy.

The massive galaxy data is consistent with a scenario where the galaxy experienced a brief epoch of high MR GC formation efficiency that was followed by a phase where the bulge enriched by $\sim60\%$ and grew by a signficant amount in total \ms.  \citet{harris_halo_2002} speculate a situation like this might occur if GCs formation tends to destroy nearby cold-gas clouds \citep[e.g. with young high-mass X-ray binaries found within GCs][]{power_primordial_2009} and hence severely limit subsequent GC formation.  Stellar bulge growth can continue from the gas expelled during the peak of GC formation, but GC formation will have ended or been severely suppressed due to the disruption of large gas clouds.  Another possibllity is that the mode of star formation transitioned from a merger-driven star formation mode to one that was characterised by less extreme pressures and/or gas turbulence \citep{elmegreen_universal_1997,ashman_constraintsformation_2001}.

While some intermediate and low-mass galaxies show metallicity offsets like those observed in massive galaxies, the typical galaxy shows similar bulge star and mean MR GC metallicities.  If this similarity reflects a coeval formation, then bulge star and MR GC formation were roughly synchronised.  

In the Milky Way, a relatively large offset in metallicity between the bulge and its MR GC subpopulation is found.  This might imply the bulge continued to grow after the GC formation epoch.

\section{The relative formation efficiency of MR GCs and bulge stars}\label{bulgeefficiency}

\begin{figure}
\center
\includegraphics[scale=0.4]{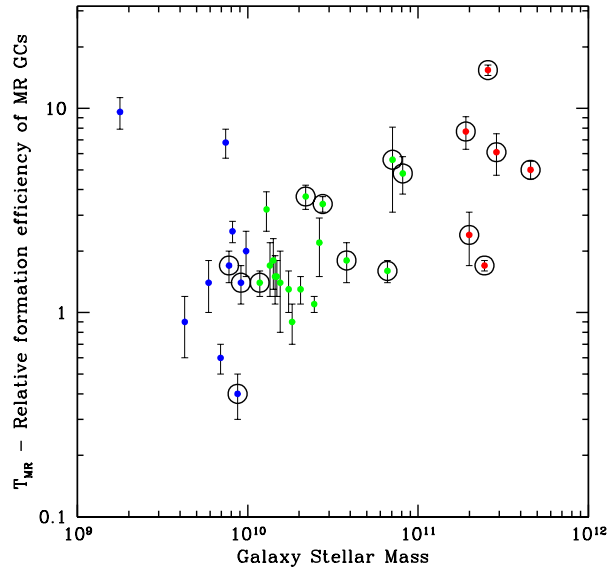}
\caption[Relative MR GC numbers against host stellar mass.]{Relative MR GC numbers (T$_{MR} = N_{GC} / (M_{stellar} / 10^9$) against host stellar mass.  Ringed solid circles have bulge stars that are more enriched by a factor of 2 compared to their MR GCs (see Fig.~\ref{figcolordiff}).  A substantial fraction of the lowest mass galaxies (\ms $\la5\times10^9$ \msun) contain no detectable MR GC subpopulation.  These are not shown in this figure, but are given in Fig.~\ref{figcolordiff}.}\label{figtred}
\endcenter
\end{figure}

In this section, the number of MR GCs normalized by its host galaxy's \ms observed today is considered.  This quantity gives some indication of the overall GC formation efficiency in a galaxy, modulo those that are destroyed through dynamical processes \citep[e.g. ][]{baumgardt_dynamical_2003}.

The relative number of MR GCs in a galaxy, when normalized by the host galaxy's stellar mass, is higher in massive galaxies compared to low and intermediate-mass galaxies \citep[e.g. ][]{rhode_metal-poor_2005,spitler_connection_2008,peng_acs_2008}.  This quantity is traditionally referred to as the MR GC T-parameter \citep[T$_{MR}$; ][]{zepf_globular_1993}. Fig.~\ref{figtred} shows T$_{MR}$ data for the ACSVCS sample.  The GC numbers were compiled as in \citet{spitler_new_2009} using the ACSVCS catalogue of \citet{peng_acs_2008}.

T$_{MR}$ parameter provides a way to gauge the efficiency of MR GC production relative to its host galaxy's {\it total} bulge \ms.  More specifically, it relates GC numbers to the bulge mass accumulated over the entire history of the galaxy.  To better understand the conditions in a galaxy {\it at the peak of GC formation}, the relative GC numbers normalized by the field stellar mass produced during {\it exactly the same epoch} should instead be considered.  This ``instantaneous'' GC formation efficiency requires knowledge of the field stars produced at the main GC formation epoch, a quantity currently unavailable for the ACSVCS galaxies.  Thus, only a qualitative assessment of the instantaneous GC formation efficiency at their peak of production can be made.

For high mass galaxies, the results of Section~\ref{mrbulgediff} suggest that the majority of a galaxy's stellar bulge formed in more enriched gas compared to gas used for MR GC formation.  If GC formation efficiency did not depend on metallicity\footnote{See \citet{forte_globular_2009} for the consequences of assuming GC formation efficiency is metallicity-dependent. Under this assumption, they are able to successfully recover a galaxy's radial and global stellar populations.} and metallicity is a timescale indicator, then the T$_{MR}$ values for such galaxies are underestimating the instantaneous GC formation efficiency.  This is because the \ms they are normalized against is much larger than the \ms produced {\em only} with the MR GCs.  Hence, the instantaneous GC formation efficiency in these galaxies will be higher than what their T$_{MR}$ parameter currently portrays.

These galaxies are highlighted in Fig.~\ref{figtred}.  The \ms in galaxies with no significant temporal bulge and MR GC metallicity offset may have formed in conjunction with the MR GC subpopulation.  Thus, their current T$_{MR}$ values may provide a reasonable qualitative picture of their inherent instantaneous GC formation efficiencies relative to massive galaxies.  Thus the overall effect of moving from the T$_{MR}$ parameter to an instantaneous GC formation efficiency will largely enhance the existing trend:  the efficiency of MR GC formation increases with galaxy \ms.

At the very low-mass end of the galaxy mass distribution, 15 out of the 25 old ACSVCS galaxies do not have significant MR GC subpopulations \citep[see ][]{forbes_bimodal_2005}, although populations of $1-3$ MR GCs likely cannot be ruled out.  This is consistent with the Local Group's 5 dwarf ellipticals (with \ms $\sim10^8$ \msun) that together host only two MR GCs \citep{sharina_ages_2006}.  If the Virgo low-mass ellipticals do not show significant MR GC subpopulations, then GC production may have been truncated due to supernova blow-out or ram-pressure stripping of its gas \citep[e.g. ][]{boselli_origin_2008}.  Low-mass galaxies also generally show high mass-to-light ratios suggesting they underwent a relatively inefficient bulge growth \citep[e.g ][]{geha_baryon_2006,brooks_origin_2007}, thus they possibly did not reach the gas densities or star-formation rates needed efficiently to produce both bulge stars and GCs.

Alternatively, if $1-3$ GCs are actually present in some of the low-mass Virgo ellipticals (this cannot be ruled out at present), the T$_{MR}$ values for these galaxies could reach levels similar to those found in massive galaxies \citep{peng_acs_2008}.  Whether this implies GC production was enhanced or the normalizing quantity in the T$_{MR}$ parameter, \ms, was underproduced cannot be determined without understanding their MR GC subpopulations, if they even exist.

\section{Cosmic metallicity enrichment}\label{redshiftmetal}

In this section, the typical formation epoch of the MR GCs is estimated using observational constraints on the early metallicity enrichment history of galaxies.  This will enable a comparison to high-redshift galaxy observations and further constrain the star-formation histories of early-type galaxies.

A new method is developed to understand the typical formation epoch of MR GCs.  By comparing the observed mean metallicity of MR GCs to empirical age-metallicity relationships (AMRs), an estimate of the typical formation epoch can be had.  GCs provide ideal objects for this method of age-dating because they are characterised as single stellar populations.  In contrast, galaxy observations are a complex mixture of multiple stellar populations, thus employing an AMR in a similar manner may not be so straight-forward.

Although this method has inherent limitations (mostly stemming from the poorly characterised AMRs available for redshifts $z\ga2$), it nevertheless provides a working approximation of the MR GC formation epoch, which can be tested with other observations or simulations.  Also, improved and/or theoretical AMRs \citep[e.g. ][]{hernquist_analytical_2003,dav_enrichment_2007,kobayashi_simulations_2007} can be easily incorporated into this analysis, which is relatively simple in design.

\subsection{Analysis}

\begin{figure*}
\center
\includegraphics[scale=0.7,angle=0]{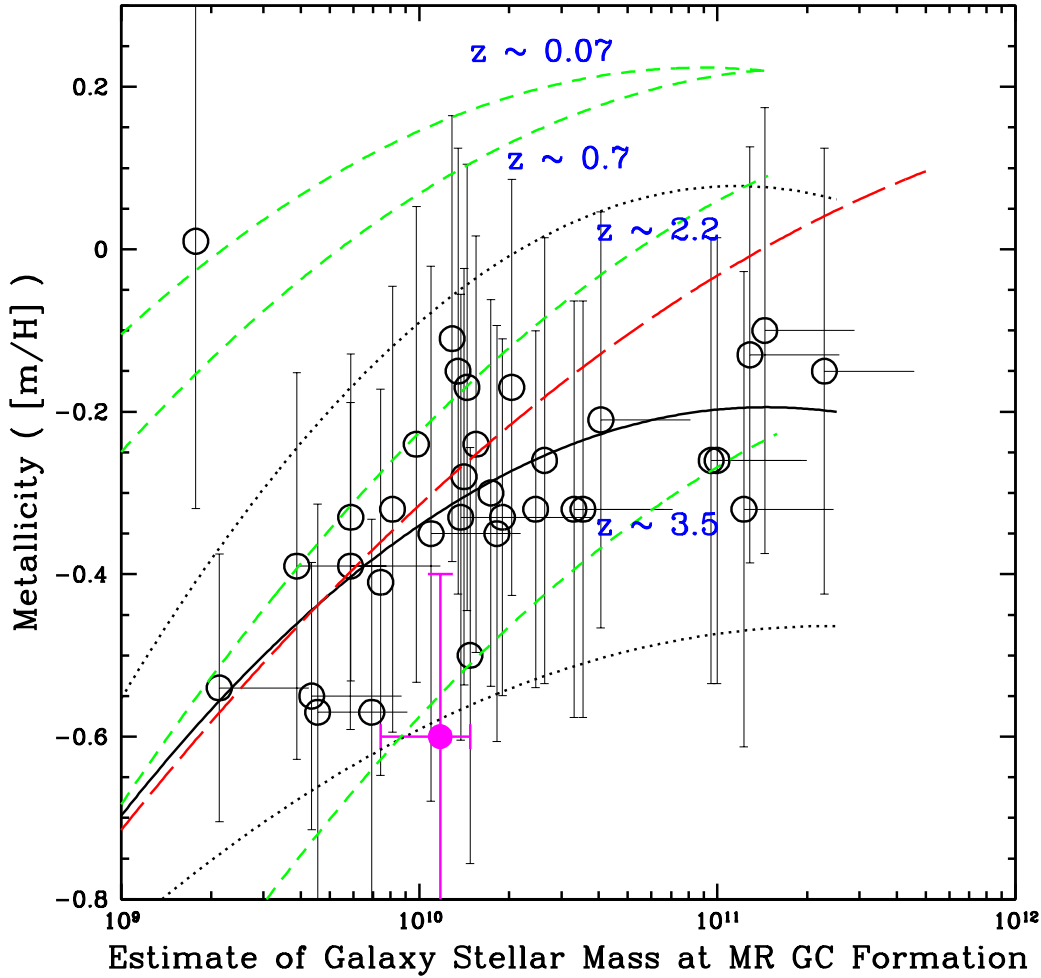}
\caption[Age-dating the formation of MR GC subpopulations.]{Age-dating the formation of MR GC subpopulations.  Observed emission-line galaxy mass-metallicity relationships (green dashed lines) for the specified redshifts \citep[compiled by ][]{maiolino_amaze._2008}.  Mean MR GC metallicities are given for Virgo early-type galaxies as open circles with a fitted curve shown as a black solid line.  Vertical error bars represent the $1\sigma$ {\it intrinsic metallicity dispersion} for the individual MR GC subpopulations, not measurement errors.  The solid, large circle represents the Milky Way MR GC subpopulation.  The relationship between the galaxy-mass metallicity relations (green dashed lines) and the fit to the corrected ACSVCS MR GCs (black solid line) suggest the MR GCs formed earlier in more massive galaxies.  The Virgo galaxy mass-metallicity relationship is given as a red, long-dashed line.  Dotted lines are fits to the $\pm1\sigma$ dispersions of the MR GC metallicity distributions.  Horizontal, one-sided errors show how much certain galaxies moved in the plot, after correcting to a \ms value that better reflects the \ms when the MR GCs were forming (see text).}\label{figmm}
\endcenter
\end{figure*}

Three empirical mass-metallicity relationships for star-forming galaxy at $z=0.07$, 0.7 and 2.2 were compiled from the literature by \citet[][ using the observations of \citealt{kewley_metallicity_2008,savaglio_gemini_2005,erb_stellar_2006}]{maiolino_amaze._2008}.  They also derived emission-line metallicity measurements in galaxies at $z\sim3.5$ and hence effectively present AMRs for a range of galaxy masses over a redshift range of $z=0-3.5$ or $\la12$ Gyrs.  \citet{mannucci_lsd:_2009} update the measurement at $z\sim3.5$ with a larger sample and use a \citet{chabrier_galactic_2003} initial mass function, both of which are adopted in the following.

Emission-line metallicities from star-forming galaxy reflect the dominant enrichment level of the gas being used in star formation.  Thus by comparing the typical MR GC metallicity to these AMRs, the peak redshift of their formation can be estimated.  Uncertainties, both intrinsic (e.g. possible environmental dependence on the AMRs) and systematic (e.g. conversion from observed gas [O/H] metallicities to stellar [m/H] metallicities; see \citealt{kobulnicky_metallicities_2004} for the conversion adopted here, assuming $Z_{\odot}=0.019$) translate into non-negligible uncertainties on the derived MR GC formation epoch.  Unfortunately, the extent of these uncertainties are not understood, thus the star formation history outlined below should be considered preliminary.  

Fig.~\ref{figmm} shows the mass-metallicity relationships for star forming galaxies at different redshifts.  To compare the ACSVCS MR GCs to these relationships, the total host galaxy \ms {\em at the time of their formation} must be known, since massive galaxies may have continued to grow in mass after the peak of MR GC formation (see \S\ref{mrbulgediff}).  As discussed in Section~\ref{bulgeefficiency}, the amount they grew after the peak of MR GC formation is currently unconstrained, so only a rudimentary correction is attempted to illustrate the corresponding implications on the general trends.  The \ms values of galaxies in Fig.~\ref{figmm} with significant bulge star and MR GC metallicity differences (offset by a factor of 2, see Fig.~\ref{figcolordiff}) have been decreased by a factor of 2.  One-sided horizontal error bars spanning this factor are provided in Fig.~\ref{figmm} for such galaxies and the quadratic fits in the Figure reflect this correction.

In Fig.~\ref{figmm}, the fit to the mean MR GC values of massive galaxies and the mass-metallicity relationships intersect at $z\sim3.5$, suggesting the peak of MR GC formation occurred in massive galaxies around this redshift.  In lower mass galaxies the intersection implies a peak of formation between $z\sim2.2-3.5$.  Thus MR GCs typically formed in massive galaxies $\sim11.7$ Gyrs ago and between $\sim10.6-11.7$ Gyrs ago in lower mass galaxies, assuming a $\Lambda$CMD cosmology and $H_0=70$ km s$^{-1}$ Mpc$^1$.  The lowest mass galaxies do not show a significant MR GC subpopulation, hence no constraint is currently available from the analysis presented in Fig.~\ref{figmm}.

The intrinsic metallicity dispersions for each MR GC subpopulation are also given as vertical error bars in the same figure.  From the intersection of the mass-metallicity relations and the upper dotted line in Fig.~\ref{figmm}, the ``end'' of the MR GC formation epoch is $z\sim2.5$ for massive galaxies and $z\sim1.5-2.0$ for lower mass galaxies.  The preferred start of the MR GC formation epoch is $z\sim4-5$ and $z\sim3.5$ for high and low mass galaxies, respectively.

As shown in Fig.~\ref{figmm}, the Milky Way's MR GC subpopulation has an implied formation age of $\sim12$ Gyrs ($z\ga3.5$), which is older than similar mass early-type galaxies in the Virgo Cluster.  Since the Milky Way's bulge appears to be more enriched than its MR GC mean (\S\ref{mrbulgediff}), the present-day \ms of the bulge may not exactly reflect its value when MR GCs were forming.  In this case, the average formation epoch may be closer to $z\sim3$.  This is still older than the cluster early-type galaxies of a similar \ms and is contrary to naive expectations for chemical enrichment histories, where denser regions of the Universe start forming stars at higher redshift.  The apparent contradiction could reflect the systematic uncertainties of employing such preliminary high-redshift galaxy mass-metallicity relationships.

\section{Discussion:  the connection to high-redshift galaxy observations}\label{disc}

\begin{figure}
\center
\includegraphics[scale=0.4,angle=0]{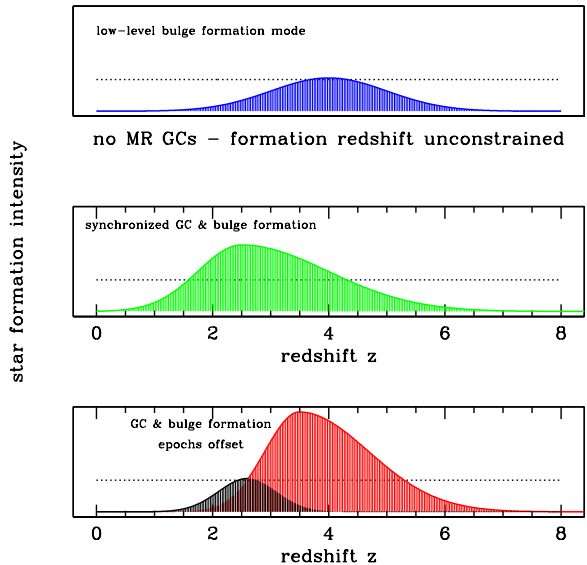}
\caption[Schematic diagram illustrating bulge formation histories derived from the analysis of MR GC subpopulations]{Schematic diagram illustrating bulge formation histories derived from the analysis of MR GC subpopulations in early-type galaxies of the Virgo Cluster.  As a rough function of the host mass (see Fig.~\ref{figmasshist}), these galaxies show three distinct bulge formation histories.  The ``star-formation intensity'' is a intentionally ambiguous term that is meant to demonstrate the fact that the exact mechanism (e.g. local star-formation rate, gas turbulence, etc.) that determines the GC formation efficiency is unknown.  The horizontal dotted line represents the threshold for GC production.}\label{figsfh}
\endcenter
\end{figure}

This section brings together the 3 topics of the preceding sections into a preliminary model of bulge star formation in Virgo early-type galaxies.  Interpretations from direct high-redshift galaxy observations are compared to this model, followed by caveats and predictions.

Fig.~\ref{figsfh} shows the implied star-formation history of early-type galaxies in Virgo as developed from the preceding analysis on their MR GC subpopulations.  The quantity on the y-axis represents the factor(s) that contribute to GC formation.  It is intentionally left ambiguous because the detailed physics of GC formation are not well-understood.  Instead the generic term ``star-formation intensity'' (SFI) is used as a proxy the mechanisms (e.g. star-formation rates, gas turbulence) that dictate GC formation efficiency within a galaxy.  Given on each panel of Fig.~\ref{figsfh} is a SFI threshold above which GC formation occurs.

The scenario in the top panel of the Fig.~\ref{figsfh} applies to very low-mass galaxies with no significant MR GC subpopulation.  Here the GC formation epochs are unconstrained and the SFI never breaches the threshold for GC formation.  Detecting strong star bursts in galaxies of \ms $\la 10^9 M_{sun}$ at redshifts $z\la4$ may prove challenging.

In the middle panel of Fig.~\ref{figsfh}, the SFI history of intermediate-mass galaxies is given.  These galaxies show similar bulge and MR GC metallicities, hence the formation epochs of the bulge stars and MR GCs are coeval.  The peak of SFI in theses galaxies is represented as a Gaussian with mean redshift of $z\sim2.5$ and spanning $z\sim1.5-4$~or~$\sim1$ Gyrs.  This age-range matches inferences made from the observed MR GC metallicity distribution spread and the observed age-metallicity relationships (see Fig.~\ref{figmm}).  The implied peak of GC and bulge star formation apparently coincides with direct observations of the cosmic star-formation density peak (at redshifts $z\sim2-3$), which some believe is dominated by star formation in low-mass galaxies \citep[e.g. ][]{hopkins_normalization_2006,bouwens_uv_2007,reddy_multiwavelength_2008}.  Here the SFI passes the GC formation threshold, to indicate GCs formed in these galaxies.  

For massive galaxies with offset bulge star and MR GC mean metallicities, the inferred SFI scenario is presented in the bottom panel of Fig.~\ref{figsfh}.  The metallicity offset implies that the bulge continued to grow after the peak of MR GC production, thus a second distribution in the SFI history is given at lower redshifts.  The extent of this bulge formation epoch is unconstrained from the present analysis, but is chosen to not overlap significantly with the MR GC formation and drop off relatively fast to match observations that a significant number of quiescent massive galaxies appear to be passively evolving at $z\sim2-3$ \citep[e.g. ][]{franx_significant_2003,van_dokkum_spectroscopic_2003,glazebrook_high_2004,daddi_passively_2005,van_dokkum_confirmation_2008,cimatti_gmass_2008,bezanson_relation_2009,naab_minor_2009}.  Although the SFI was low and few MR GCs formed, the bulge likely doubled in total \ms and its global metallicity increased by a factor of 2 during this second mode of bulge growth.  The main GC SFI distribution reaches very high values of SFI to signify the ultra-efficient period of MR GC production supported by the observation (see Fig.~\ref{figtred}).  The GC and bulge epochs overlap to illustrate the fact that some very enriched MR GCs exist.

In massive galaxies, the peak of MR GC formation occurred at $z\sim3-4$.  This redshift is somewhat earlier than the observed peak in the number density of submillimeter galaxies \citep[$z\sim2.4$; ][]{chapman_redshift_2005}, whose properties can be explained by very intense star-formation rates of $\sim1000$ \msun~yr$^{-1}$ \citep{hughes_high-redshift_1998,blain_submillimeter_2002,swinbank_properties_2008}.  These strong star formation events were possibly induced by major galaxy mergers \citep{tacconi_submillimeter_2008} or disk instabilities \citep[e.g. ][]{escala_stability_2008,shapiro_star-forming_2010} from rapid, cold-gas accretion \citep{dekel_cold_2009}.  In the model developed here, this epoch is followed by a mode of bulge growth that is not as efficient at GC production.  Observationally, this period may manifest itself as the heavily dust-obscured massive galaxies found at $z\sim2$ \citep[e.g. ][]{dey_significant_2008,bussmann_hubble_2009}, which just finished a submillimeter-bright phase \citep[see discussion in ][]{dey_significant_2008}.

At roughly the same time, approximately $z\sim2.5$ or $\sim11$ Gyrs ago, intermediate-mass galaxies were experiencing their peak of bulge growth {\it and} MR GC formation.  The implied SFIs may resemble those in massive galaxies at the same epoch.  The average age of bulges in both massive and intermediate-mass galaxies should therefore be roughly identical, despite a relatively large difference in their mean metallicities.

\subsection{Caveats}

The above scenario depends on the following:
\begin{itemize}
\item That there is no significant age difference between the galaxy bulge stars and MR GCs.  Even though obviously young galaxies are removed from the analysis, transforming galaxy broadband colours into metallicities can be complicated by undetected age variations between the galaxies and MR GCs.
\item The assumption that the destruction of GCs, be it from disrupting tidal forces of the host galaxy or otherwise, does not influence the above results significantly.
\item The use of uncertain galaxy age-metallicity relationships from high-redshift observations.  Environmental and morphological differences could translate into large systematic uncertainties on the attempt to age-date the peak of GC formation.
\item The assumption that the apparent offsets in bulge star and MR GC metallicities translates into an offset in time.  Gas accretion, mixing and outflows contribute to inhomogeneities in the spatial metallicity distributions of a galaxy and will all sabotage the simple model presented here.
\end{itemize}
The last two points can be better understood by incorporating detailed predictions from theoretic simulations of cosmological metal-enrichment and galaxy formation \citep[e.g. ][]{hernquist_analytical_2003,dav_enrichment_2007,kobayashi_simulations_2007}.  These issues will also benefit from more detailed observational analysis of galaxies at high-redshift \citep[e.g. ][]{calura_evolution_2009}.

\subsection{Predictions}

Presented below are a number of predictions resulting from the model outlined above.

\begin{figure}
\center
\includegraphics[scale=0.4]{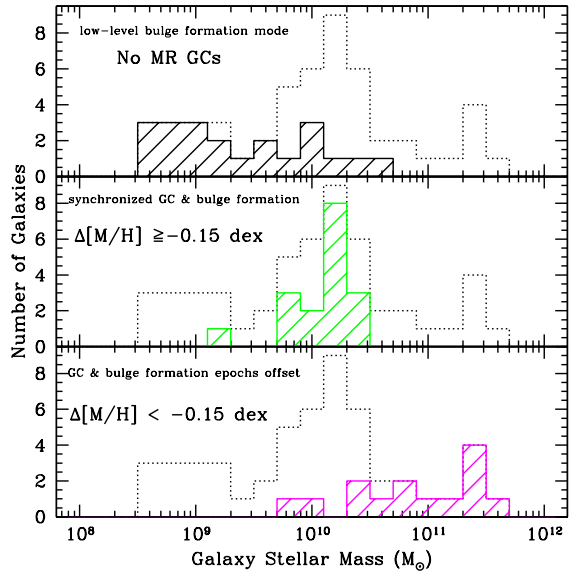}
\caption[Galaxy \ms histograms for old ACSVCS early-type galaxies]{Galaxy \ms histograms for old ACSVCS early-type galaxies.  Dotted histograms show the total \ms distribution of the sample.  Solid histograms in upper, middle and lower panel correspond to galaxies with no MR GCs, with no significant and a significant offset between the typical enrichment of its MR GCs and its bulge, respectively.  These are classified according to the observed metallicity offset between their MR GCs and bulge stars, as shown in Fig.~\ref{figcolordiff}.  At \ms $\sim10^{10}$ \msun, the 3 classes overlap.}\label{figmasshist}
\endcenter
\end{figure}

{\bf Understanding the mode of bulge star formation around the transition galaxy stellar bulge mass.} The three bulge formation histories presented in Fig.~\ref{figsfh} roughly correlate with galaxy \ms, as shown in Fig.~\ref{figmasshist}.  At \ms $\sim10^{10}$ \msun the three modes overlap.  This bulge \ms corresponds to the characteristic mass where many galaxy properties transition \citep[e.g. ][]{dekel_galaxy_2006}.  This work presents predictions for the bulge star formation histories of galaxies below, at and above this transition bulge \ms.

{\bf GC and galaxy ages.} The derived MR GC formation redshifts in massive and intermediate-mass galaxies suggests an age offset of $\sim1$ Gyr should exist in their MR GC subpopulations.  Typical theoretical modelling and measurement uncertainties mean that absolute ages of individual extragalactic GCs are difficult to constrain with such precision.  However, in principle, two very large samples of at least $1000$ MR GC ages (one for each galaxy-mass class) derived from standard Lick absorption-line analysis can be used to determine whether or not such a {\it relative} age offset exists.  This minimum sample size is derived assuming a mean MR GC age of 11.0 Gyrs (or log age $=1.04$ in Gyrs) and 11.7 Gyrs (log age $=1.07$) in intermediate and massive galaxies, respectively.  Typical uncertainties are assumed to be $\pm0.15$ on the logarithm of the age (in Gyrs) of individual GCs \citep[e.g. ][]{proctor_keck_2008}.  Note, systematic uncertainties from stellar population models and the age-metallicity degeneracy may complicate such attempts.  Furthermore, the spread of metallicity may imply a spread of age, thus making this prediction more challenging to measure directly.

Another interesting test of this model is to age-date the bulge stars of a massive galaxy (such as Virgo's Messier~87) and determine whether the mean age of 1000s of its MR GCs are offset from the bulge's average age by $\sim1$ Gyr.  A similar offset is expected in the Milky Way.

{\bf GC formation efficiency.}  The discovery of an offset between the mean metallicity of MR GCs and their host bulge could mean that a significant fraction of the bulge \ms in massive galaxies formed after their typical MR GCs.  If the relative MR GC numbers observed in galaxies (see Fig.~\ref{figtred}) were instead normalized by only the galaxy \ms that formed in conjunction with the MR GCs and it is assumed that GC formation efficiency does not depend on metallicity, an extremely MR GC efficient formation in massive galaxies is implied.  The apparent prevalence of ultra-compact dwarfs (which are sometimes considered as massive MR GCs; e.g. see \citealt{forbes_uniting_2008}) around massive galaxies may be a natural by-product of an ultra-efficient GC formation epoch \citep[see discussion in ][]{larsen_mass_2009}.  This would mean that the local star-formation intensity history, not environment, dictates where such massive compact star clusters will be found \citep{hau_ultra-compact_2009}.

\section*{Acknowledgments}

The anonymous referee provided insightful comments that greatly improved the manuscript.  We thank D. Forbes, A. Romanowsky and J. Strader for reviewing a draft of the manuscript and are grateful for useful discussions with A. Alves-Brito.  LS appreciates the hospitality of S\o ren Larsen and others at Utrecht University, where part of this work was written.


\label{lastpage}

\end{document}